# Trade-based Asset Model using Dynamic Junction Tree for Combinatorial Prediction Markets

WEI SUN, KATHRYN B. LASKEY, CHARLES TWARDY, ROBIN HANSON, George Mason Univ.
BRANDON GOLDFEDDER, Gold Brand Software

## 1. INTRODUCTION

Prediction markets have demonstrated their value for aggregating collective expertise [Arrow et al. 2008]. Combinatorial prediction markets allow forecasts not only on base events, but also on conditional and/or Boolean combinations of events [Hanson 2007]. We describe a trade-based combinatorial prediction market asset management system, called Dynamic Asset Cluster (DAC), that improves both time and space efficiency over the method of Sun et al. [2012], which maintains parallel junction trees for assets and probabilities. The basic data structure is the asset block, which compactly represents a set of trades made by a user. A user's asset model consists of a set of asset blocks representing the user's entire trade history. A junction tree is created dynamically from the asset blocks to compute a user's minimum and expected assets.

## 2. DYNAMIC ASSET CLUSTER MODEL

Unless otherwise stated, capital letters stand for random variables, lowercase letters stand for states of random variables, and bold letters stand for sets. The global joint state is indexed by $i$; $p_i$ denotes current probability of the state $i$; $x_i$ is the new trade/edit on state $i$; $b$ is a market thickness/scaling parameter; and $a$ denotes the user's assets. Superscripts are used for user's trade index: e.g., $a_i^0$ represents the initial assets at state $i$, while $a_i^1$ stands for a user's assets at state $i$ after the first trade.

### 2.1 Trade-based Asset Block

For each individual trade, DAC constructs a new asset block and/or updates existing asset blocks. We define asset block as follows:

*Definition* 2.1 (*Asset Block*). For a general trade $x(T|\mathbf{U} = \mathbf{u})$, where $T$ is the target question, $\mathbf{U}$ is a set of assumed questions, and $\mathbf{u}$ is the assumed state of $\mathbf{U}$, the trade-based **asset block** is the joint space of $T, \mathbf{U}$ associated with an asset table containing asset values for all joint states of $T, \mathbf{U}$.

Letting $\mathbf{W}$ denote all other questions in the market, we know from the probability chain rule that

$$\begin{aligned} x_i &= x(\mathbf{W}|T, \mathbf{U})x(T|\mathbf{U} = \mathbf{u})x(\mathbf{U} = \mathbf{u}) \\ &= p(\mathbf{W}|T, \mathbf{U})x(T|\mathbf{U} = \mathbf{u})p(\mathbf{U} = \mathbf{u}) \end{aligned}$$

Therefore,

$$\frac{x_i}{p_i} = \frac{x(T|\mathbf{U} = \mathbf{u})}{p(T|\mathbf{U} = \mathbf{u})}$$

If this is the first trade, then after the trade $x(T|\mathbf{U} = \mathbf{u})$, according to the market scoring rule [Hanson 2007], the user's assets become

$$a_i^1 = a_i^0 + b\log(\frac{x_i}{p_i})$$

$$= a_i^0 + b\log(\frac{x(T|\mathbf{U}=\mathbf{u})}{p(T|\mathbf{U}=\mathbf{u})})$$

$$\prec b\log(\frac{x(T|\mathbf{U}=\mathbf{u})}{p(T|\mathbf{U}=\mathbf{u})})$$

The $\prec$ in the third row above (meaning 'determined by') is because the initial asset value is arbitrary.

Now the key question is how to assemble the user's assets when we have more trades by the user. Suppose the user has made another trade $x(E|\mathbf{R}=\mathbf{r})$. The following lists four exhaustive and collective scenarios for the new trade:

(1) $\{E,\mathbf{R}\}$ is a subset of $\{T,\mathbf{U}\}$ (including the case where $\{E,\mathbf{R}\}$ is exactly the same as $\{T,\mathbf{U}\}$) - update the asset block of $T,\mathbf{U}$ with the new trade.
(2) $\{E,\mathbf{R}\}$ is a superset of $\{T,\mathbf{U}\}$ - create a new asset block of $\{E,\mathbf{R}\}$, and then merge $\{T,\mathbf{U}\}$ into $\{E,\mathbf{R}\}$.
(3) $\{E,\mathbf{R}\}$ is a completely separate set of variables from $\{T,\mathbf{U}\}$ - create a new asset block.
(4) $\{E,\mathbf{R}\}$ has overlapped variables (intersection set is denoted as $\mathbf{K}$) with $\{T,\mathbf{U}\}$, and $\{E,\mathbf{R}\}$ also has variables that are not in $\{T,\mathbf{U}\}$ - create a new asset block.

Note: when we merge asset blocks, we must ensure that only one existing block is merged into the new one; otherwise it could result in double counting. After merging, we need to make sure the old block is deleted.

The collection of a user's asset blocks is a compact representation of the user's trading history. We can safely calculate all asset related values by iterating over all possible permutations of states in the joint asset blocks space. But the brute force iterating method is not scalable when number of overlapping variables between asset blocks becomes big. Fortunately, we can do better by utilizing an asset junction tree.

2.2 Computing Minimum Asset & Expected Asset by Asset Junction Tree

Variables in any individual asset block are viewed as related and so can be pair wise connected to comprise an undirected graph. Graph theory tells that triangulation applied to the undirected graph can find all strongly connected components. The cliques of the graph can be assembled into a junction tree whose nodes are the cliques. Each clique is made up of a set of strongly connected components from the original graph. The asset table for each clique is constructed by merging those asset blocks which are subsets of the clique space. The derivation of the graph transformation can be found in [Pearl 1988].

The traditional junction tree algorithm uses two-way propagation to obtain the correct joint probability for each clique. In the combinatorial prediction market context, we need to find the minimum joint asset because the user's assets must never be allowed to become negative. The minimum assets can be computed by customized one-way propagation as shown below:

**One-way min-propagation**: After choosing any clique to be a root, we can then determine a one-way propagation order from all leaf nodes to the chosen root. Now the problem is how to propagate assets from one clique to its neighbor. Assume that two cliques $\mathbf{C}_i$ and $\mathbf{C}_j$ are neighbors in the junction tree, and the separator $\mathbf{S}_{ij}$ is associated with the edge between $\mathbf{C}_i$ and $\mathbf{C}_j$. The asset tables for $\mathbf{C}_i, \mathbf{C}_j$ and $\mathbf{S}_{ij}$ are $\phi(\mathbf{C}_i), \phi(\mathbf{C}_j)$ and $\phi(\mathbf{S}_{ij})$ respectively. Min-propagation from $\mathbf{C}_i$ to $\mathbf{C}_j$ along the separator $\mathbf{S}_{ij}$ follows the min-propagation protocol presented below.

(1) Let $\phi(\mathbf{S}_{ij})' = \min_{\mathbf{C}_i \setminus \mathbf{S}_{ij}} \phi(\mathbf{C}_i)$, — minimizing $\phi(\mathbf{C}_i)$ onto the domain of the separator $\phi(\mathbf{S}_{ij})$.

(2) Let $\mathcal{L}(\mathbf{S}_{ij}) = \phi(\mathbf{S}_{ij})' - \phi(\mathbf{S}_{ij})$, — subtracting the old asset in the separator $\mathbf{S}_{ij}$ with the projected minimization value from its neighbor. The result is called the separator gain.

(3) Let $\phi(\mathbf{C}_j) = \phi(\mathbf{C}_j) + \mathcal{L}(\mathbf{S}_{ij})$, — summing up with the separator gain to update the assets $\phi(\mathbf{C}_j)$.

Computing expected assets is straightforward once the inference engine provides joint probabilities for the clique structure in the asset junction tree. We simply multiply assets times corresponding probabilities then sum.

### 2.3 Update asset block after resolving questions

When we resolve a question from the market, first we need to update the market distribution given the settlement. We then need to update all users' asset blocks. Basically, resolving a question for asset block is to instantiate a particular dimension then reduce the associated asset table. In the extreme case when an asset block has only one question which is the one being resolved, the asset block collapses into one scalar number after resolving.

### 3. COMPUTATIONAL EXAMPLES & PERFORMANCE

We implemented DAC model and compared its performance in computing user's cash with a brute force method that iterates over asset blocks. As expected, DAC performs very well when there are many overlapping variables across blocks because it utilizes the efficiency of tree propagation, while the brute force method performs a little better when the asset table is sparse and blocks do not overlap much. Figure 1 (a) demonstrates the time cost comparison in a scenario that consists of a chain of trades. Figure 1 (b) shows a time comparison when there are only two asset blocks but the size of block increases. Note the graphs are logarithmic: on balance, the results favor DAC.

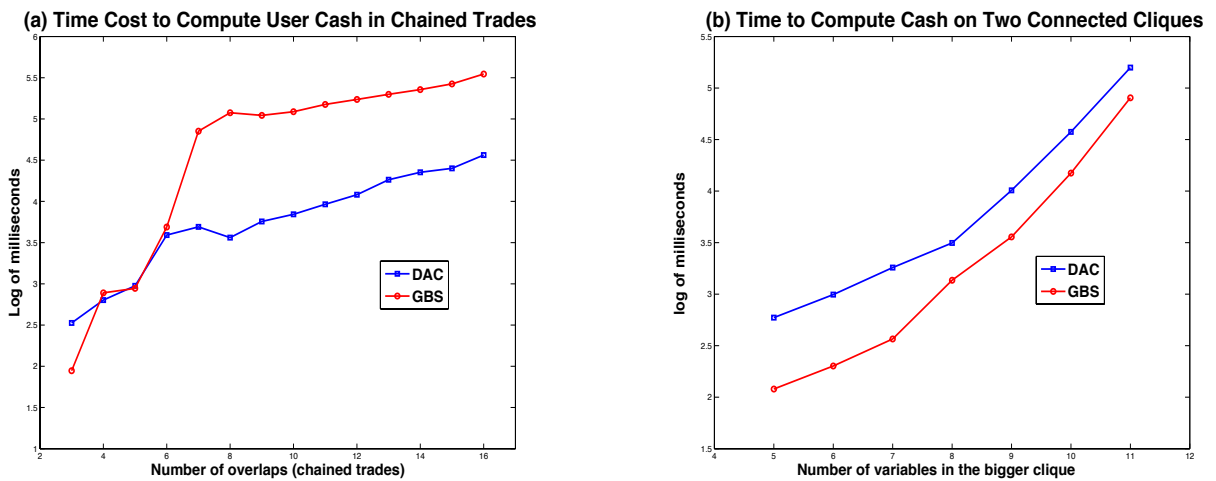

Fig. 1. Time cost comparison between DAC and a brute force iterating implementation called GBS

ACKNOWLEDGMENTS
This research was supported by the Intelligence Advanced Research Projects Activity (IARPA) via Department of Interior National Business Center contract number D11PC20062. The U.S. Government is authorized to reproduce and distribute reprints for Governmental purposes notwithstanding any copyright annotation thereon. Disclaimer: The views and conclusions contained herein are those of the authors and should not be interpreted as necessarily representing the official policies or endorsements, either expressed or implied, of IARPA, DoI/NBC, or the U.S. Government.

REFERENCES
Kenneth Arrow, Robert Forsythe, Michael Gorham, Robert Hahn, Robin Hanson, John O. Ledyard, Saul Levmore, Robert Litan, Paul Milgrom, Forrest D. Nelson, George R. Neumann, Marco Ottaviani, Thomas C. Schelling, Robert J. Shiller, Vernon L. Smith, Erik Snowberg, Cass R. Sunstein, Paul C. Tetlock, Philip E. Tetlock, Hal R. Varian, Justin Wolfers, and Eric Zitzewitz. 2008. The Promise of Prediction Markets. *Science* 320, 5878 (May 2008), 877–878. DOI:http://dx.doi.org/10.1126/science.1157679

Robin Hanson. 2007. Logarithmic Market Scoring Rules for Modular Combinatorial Information Aggregation. *Journal of Prediction Markets* 1, 1 (2007), 3–15.

Judea Pearl. 1988. *Probabilistic Reasoning in Intelligent Systems: Networks of Plausible Inference* (1st. ed.). Morgan Kaufmann.

Wei Sun, Robin Hanson, Kathryn B. Laskey, and Charles Twardy. 2012. Probability and Asset Updating using Bayesian Networks for Combinatorial Prediction Markets. In *Proceedings of the 28th Conference on Uncertainty in Artificial Intelligence (UAI-12)*. Catalina, CA.